\documentclass[structabstract]{aa}  
\usepackage{graphicx}
\usepackage{txfonts}
\usepackage{longtable}
\usepackage[authoryear]{natbib}
\bibpunct[; ]{(}{)}{;}{a}{}{,}

\newcommand{\n}{\'{n}}

\newcommand{\obj}{\object}

\begin{document}
   \title{Orbital and physical parameters of eclipsing binaries from the ASAS catalogue}
   \subtitle{III. Two new low-mass systems with rapidly evolving spots.}
  
   \titlerunning{Two low-mass systems with evolving spots}

   \author{K. G. He\l miniak\inst{1,2} \and M. Konacki\inst{1,3} \and 
	K. Z\l oczewski\inst{4} \and M. Ratajczak\inst{1} \and D. E. Reichart\inst{5}
	\and K. M. Ivarsen\inst{5} \and J. B. Haislip\inst{5} \and J.~A.~Crain\inst{5}
	\and A. C. Foster\inst{5} \and M. C. Nysewander\inst{5} 
	\and A. P. LaCluyze\inst{5}
          }

   \institute{Nicolaus Copernicus Astronomical Center, Department of Astrophysics,
              ul. Rabia\'nska 8, 87-100 Toru\'n, Poland\\
              \email{xysiek@ncac.torun.pl}
	 \and
	     Departamento de Astronom\'{i}a y Astrof\'{i}sica, Pontificia Universidad 
		Cat\'{o}lica, Vicu\~{n}a Mackenna 4860, Casilla 306, Santiago 22, Chile
         \and
             Astronomical Observatory, Adam Mickiewicz University, ul. S\l oneczna 36,
		60-286 Pozna\'n, Poland
	 \and
	     Nicolaus Copernicus Astronomical Center, ul. Bartycka 18, 00-716 Warszawa,
		Poland
	\and
	Department of Physics and Astronomy, University of North Carolina, 
	Campus Boc 3255, Chapel Hill, NC 27599-3255, USA\\
             }

   \date{Received ...; accepted ...}

 
  \abstract
   {}
   {We present the results of our spectroscopic and photometric analysis
of two newly discovered low-mass detached eclipsing binaries found in
the \emph{All-Sky Automated Survey} (ASAS) catalogue: 
\object{ASAS J093814-0104.4} and \object{ASAS J212954-5620.1}.}
   {Using the Grating Instrument for Radiation Analysis with a Fibre-Fed Echelle
(GIRAFFE) on the 1.9-m Radcliffe telescope at the South African Astronomical Observatory 
(SAAO) and the University College London Echelle Spectrograph (UCLES)
on the 3.9-m Anglo-Australian Telescope, we obtained high-resolution spectra of 
both objects and derived their radial velocities (RVs) at various orbital phases. 
The RVs of both objects were measured with the two-dimensional cross-correlation 
technique (TODCOR) using synthetic template spectra as references. We also obtained 
$V$ and $I$ band photometry using the 1.0-m Elizabeth telescope at SAAO and
the 0.4-m Panchromatic Robotic Optical Monitoring and Polarimetry
Telescopes (PROMPT) located at the Cerro Tololo Inter-American Observatory 
(CTIO). The orbital and physical parameters of the systems were derived with 
PHOEBE and JKTEBOP codes. We compared our results with several sets of 
widely-used isochrones.}
   {Our multi-epoch photometric observations demonstrate that both objects show 
significant out-of-eclipse modulations, which vary in time. We
believe that this effect is caused by stellar spots, which
evolve on time scales of tens of days. For this reason, we constructed our 
models on the basis of photometric observations spanning short time scales 
(less than a month). Our modeling indicates that 
(1) \object{ASAS J093814-0104.04} is a main sequence
active system with nearly-twin components with masses of 
$M_1 = 0.771 \pm 0.033$ M$_{\odot}$, $M_2 = 0.768 \pm 0.021$ M$_{\odot}$ and radii
of $R_1 = 0.772 \pm 0.012$ R$_{\odot}$ and $R_2 = 0.769 \pm 0.013$ R$_{\odot}$.
(2) \object{ASAS J212954-5620.1} is a main sequence active binary with
component masses of $M_1 = 0.833 \pm 0.017$ M$_{\odot}$, 
$M_2 = 0.703 \pm 0.013$ M$_{\odot}$ and radii of 
$R_1 = 0.845 \pm 0.012$ R$_{\odot}$ and $R_2 = 0.718 \pm 0.017$ R$_{\odot}$.}
   {Both systems seem to confirm the well-known
characteristic of active low-mass stars, for which the observed radii are larger and
the temperatures lower than predicted by evolutionary models. Other parameters
agree within errors with the models of main sequence stars. The 
time-varying spot configuration may imply a variable level of activity, which
may manifest itself in small changes of the measured radii.}


   \maketitle
%

\section{Introduction}

Since the discovery of the first eclipsing late-type systems -- \obj{YY~Gem}
\citep{kro52,bop74}, \obj{CM~Dra} \citep{lac77} or \obj{CU~Cnc} 
\citep{del99} -- and the first comparisons of their observed properties 
with the evolutionary models \citep[e.g.][]{lac77}, it became clear that 
the stellar evolution theory, despite being very successful in many 
fields, fails to reproduce the observed parameters of late-type stars, 
especially radii and temperatures. An increasing number of recently 
discovered systems (see \citealt{tor10} for a recent review) confirms 
these discrepancies at least for systems on the main sequence. 
Predicted radii of low-mass stars are systematically larger and temperatures 
lower than the ones derived from observations. It is now claimed 
that this is caused by the improper treatment of stellar activity and 
convection in outer layers of the stellar interior. A presence of a close 
companion and its gravitational influence seems to increase the
activity level and suppresses the effectiveness of convection. 
Considering that the amount of energy produced in the core is the same 
for single stars as for close binary components, this leads to larger radii 
and lower effective temperatures while at the same time keeping the overall 
luminosity at an unchanged level \citep{cha07,mor09}. However, in order
to better constrain and understand these differences between models and 
observation, more low-mass systems showing various levels of activity and
with accurately known parameters are needed.

In this paper we present the first analysis together with the orbital 
and physical parameters of two new spectroscopic detached eclipsing 
binaries (DEBs), both with component masses below 1 M$_{\odot}$ -- 
\object{ASAS J093814-0104.4} (hereafter ASAS-09) and 
\object{ASAS J212954-5620.1} (hereafter ASAS-21) -- and both
showing substantial level of activity. For ASAS-09 we reach 
2.7~-~4.3~\% level of precision in the masses and $\sim$ 1.6 \%
in radii. This is better or close to the level of 3 \% required for stringent 
tests of the evolutionary models \citep{bla08,cla08}. For ASAS-21
we reach $\sim$ 2.0 \% for masses and 1.4~-~2.4~\% for radii.
Both systems were noticed to present significant out-of-eclipse
and time-varying brightness modulations, which we deduce to be 
caused by spots and their evolution on the time scale
of single weeks. 

Both objects are be interesting from the evolutionary point 
of view. ASAS-09 is composed of two nearly identical dwarfs and 
iso be a good candidate for testing the so-called Vogt or
Vogt-Russel theorem \citep{vog26}. ASAS-21 on the contrary shows
one of the lowest (still quite high however) mass ratios measured 
in low-mass binaries  -- $q\sim0.85$. The components of systems with 
low $q$ lay relatively far from each other on the isochrones,
which allows to perform more stringent tests of the stellar models 
than for systems with $q$ close to 1.

\section{ASAS-09 and ASAS-21}

\subsection{ASAS-09 (\object{GSC 04894-02310})}
The eclipsing binary \object{ASAS J093814-0104.4} is the faintest of
all low-mass detached eclipsing binaries in our sample of systems found in
the ASAS catalogue and has $V = 12.07 \pm 0.12$ mag \citep{poj02}.
Its eclipsing nature was noted for the first time in 
the \emph{ASAS Catalogue of Variable Stars}\footnote{http://www.astrouw.edu.pl/asas/} 
\citep[ACVS; ][]{poj02}. The system is classified in ACVS as an 
eclipsing detached binary (ED) with the amplitude of the photometric 
variation of $0.47$ mag and $(V - I) = 1.12$ mag \citep{szc08}. It can 
be associated with a soft X-ray source \object{1RXS~J093813.2-010423}
\citep[$F_X = 0.40 \pm 0.06 \times 10^{-12}$ erg cm$^{-2}$ s$^{-1}$;][]{vog99}

The ASAS data phase coverage is full and the time span covers almost nine
years but, the scatter reaches almost 0.4 mag not only due to the quality of
the data but also due to the evolution of stellar spots on relatively 
short time scales. The ASAS light curve phase-folded with the orbital period
from our analysis is presented in Figure \ref{fig_lc_asas}. It shows nearly 
equal eclipses and, not surprisingly, no clear out-of-eclipse variations.

\subsection{ASAS-21 (\object{GSC 08814-01026})}
The eclipsing system \object{ASAS J212954-5620.1} has an apparent $V$ 
magnitude of $11.83 \pm 0.10$ \citep{poj02}.
The first information about its variability or binarity can be found in 
the ACVS, where this system is classified as ED with the amplitude of 
the photometric variation of 0.77 mag and $(V-I)=1.29$. It also has 
a soft X-ray identification -- \object{1RXS J212954.0-561954} 
($F_X=0.714\pm0.102 \times 10^{-12}$ erg cm$^{-2}$ s$^{-1}$).

\begin{figure*}
\centering
\includegraphics{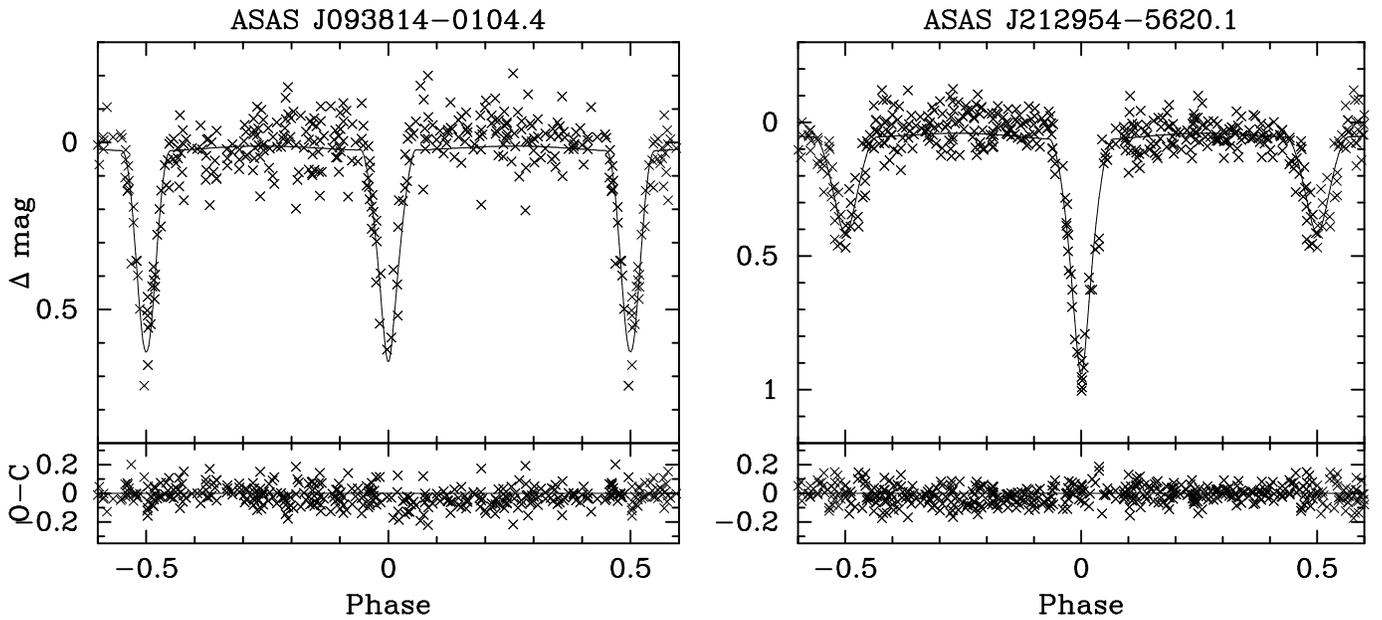}
\caption{ASAS $V$ band light curves of ASAS-09 (left) and ASAS-21 (right)
phase-folded with the best fitting orbital periods (see Table \ref{tab_orb}).
Solid lines represent models based on our solutions (see Section \ref{sect_phot}) 
but without spots.}\label{fig_lc_asas}
\end{figure*}

As for the previous system, the available ASAS photometry spans about nine
years and covers the entire orbit. The ASAS curve phase-folded with 
the period obtained from our modeling is shown in Fig. \ref{fig_lc_asas}.
One can see a considerable difference in the depths of eclipses and hints 
of an out-of-eclipse variation. The photometric scatter is increased
by the evolution of the spots as well.

\section{Observations}

\begin{figure*}
\centering
\includegraphics{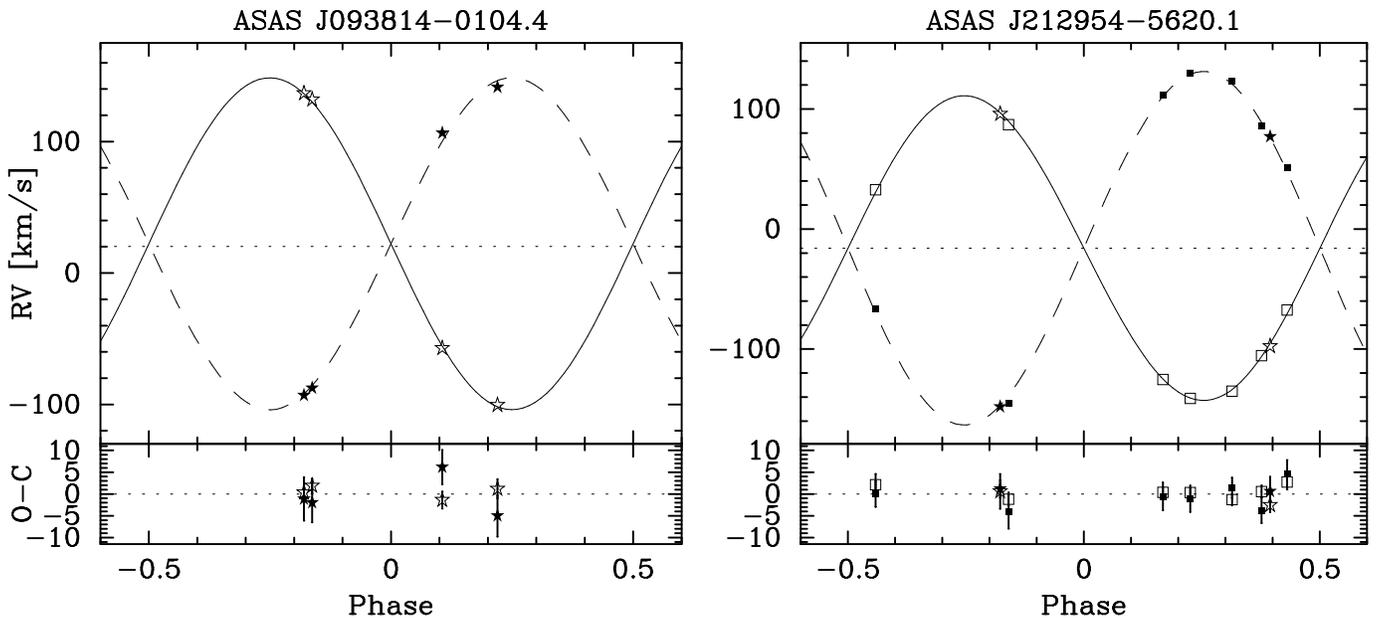}
\caption{Radial velocities of ASAS-09 (left) and ASAS-21 (right) from
	our AAT/UCLES (stars) and Radcliffe/GIRAFFE (squares) spectra 
	together with the best fitting models. The measurements for the
	primaries are plotted with open and for the secondaries with filled symbols.
	The solid line shows the best fit for the primaries and the dashed line for 
	the secondaries. The dotted line shows the systemic velocity $\gamma$.
	\label{fig_rv}}
\end{figure*}

\subsection{Spectroscopy}

ASAS-09 and ASAS-21 were observed spectroscopically as candidate
late-type detached eclipsing binaries (DEBs) in order to obtain their
radial velocity (RV) curves and confirm their low masses. The latter system was 
observed earlier, at the very beginning of our survey, during four
consecutive nights in June 2006. We used the SAAO's 1.9-m Radcliffe telescope and
its Grating Instrument for Radiation Analysis with a Fibre-Fed Echelle
(GIRAFFE) to obtain a series of R$\sim$40000 spectra. Because of the 
faintness of the target we set the exposure time to 3600 seconds. The spectra 
were later binned four times in the dispersion direction in order to
improve their low signal-to-noise ratio (SNR). In the best cases, we reached 
SNR$\sim$70 per collapsed pixel in the center of an \'echelle order near
$\lambda = 6500$ \AA. This value varied with weather conditions (seeing).
The wavelength calibration was done in the standard manner with the ThAr lamp 
exposures taken before and after a stellar exposure.

Two additional spectra of ASAS-21 were taken with the AAT/UCLES during our
observing run in September 2008. We used a 1 arcsec slit, which provides a 
resolution of R$\sim$60000. With 900 seconds exposures and no binning we 
reached a significantly lower SNR$\sim$25, but a comparable RV precision
of several km~s$^{-1}$.

We observed ASAS-09 in January 2009 with the same configuration of 
the AAT/UCLES. Despite even lower SNR ($\sim15-20$) we were able to 
derive RV measurements with a km~s$^{-1}$ precision. 
As in the case of Radcliffe/GIRAFFE data,
the wavelength calibration was done in the standard manner with the ThAr 
lamp. An additional drawback during our Australian observations was 
a very poor seeing (typically well above 2 arcsec). However, we were still 
able to obtain useful data for the two objects.

The radial velocities for both systems were obtained in the same way as
described in our previous papers  -- \citet[][hereafter paper~1]{hel09} 
and \citet[][hereafter paper~2]{hel10} -- with the TODCOR technique 
\citep{zuc94} and the synthetic ATLAS spectra \citep{cas03} as templates. 
The iodine cell available at AAT/UCLES was not used because of the 
low brightness of both targets. Owing to the low SNR and the
rotational broadening, we did not perform any abundance analysis.

\subsection{Photometry\label{sect_phot}}

\begin{table}
\caption{Number of photometric data points for each data set in the
	$V$ and $I$ bands for both systems.}\label{tab_data_no}
\centering
\begin{tabular}{lcc|lcc}
\hline \hline
 &{\bf ASAS-09}& & & {\bf ASAS-21} & \\
Data set & $V$ band & $I$ band & Data set & $V$ band & $I$ band \\
\hline
 2008-01 &  41 &  43 & 2008-05 & 166 & 164 \\
 2008-12 &  97 & 113 & 2009-06 & 127 & 159 \\
 2009-03 & 140 & 159 & 2009-07 & 120 & 120 \\
 2010-01 & 232 & 292 & 2009-10 & 129 & 118 \\
\hline
\end{tabular}
\end{table}

Because ASAS-09 is a short period and a well detached system, it 
was considered as a low-mass candidate from the beginning of our 
survey. Thus it was observed photometrically well before the 
spectroscopic follow-up. We used the 1.0-m Elizabeth telescope at 
the South African Astronomical Observatory (SAAO) and a 
$1024 \times 1024$ STE4 camera with the Johnson $V$ and
Cousins $I$ filters. The field of view was $317 \times 317$ arcseconds
(0.31 arcsec/pix) and the exposure time varied in order to compensate 
for varying observing conditions. The data were processed with 
the standard data reduction tasks available in the \emph{IRAF} 
package\footnote{\emph{IRAF} is written and supported by the 
\emph{IRAF} programming group at the National 
Optical Astronomy Observatories (NOAO) in Tucson, AZ. NOAO is 
operated by the Association of Universities for Research in 
Astronomy (AURA), Inc. under cooperative agreement with the 
National Science Foundation. http://iraf.noao.edu/}.

Three separate data sets, covering both eclipses (without 
the secondary minimum) and their vast surroundings, were obtained in 
January 2008, December 2008, and March 2009. A clear out-of-eclipse
modulation is seen, which originates in stellar spots. 
However, we failed to produce a final model of the system, because
the data sets occurred to be inconsistent, which can be explained by an
evolution of the spot pattern. Considering this, we decided to carry out 
additional observations using the Panchromatic Robotic Optical 
Monitoring and Polarimetry Telescopes (PROMPT).

PROMPT, located at CTIO in Chile, consists of six robotic 0.41-m 
Ritchey-Chr\'{e}tien telescopes equipped with rapid-res\-ponse Alta~U47+ 
cameras and a wide range of astronomical filters. Telescopes serve 
mainly for taking rapid and simultaneous multi-wavelength observations 
of gamma-ray bursts (GRB) afterglows. PROMPT is operated by SKYNET -- 
a distributed network of robotic telescopes located around the world, 
dedicated for continued GRB afterglow observations. For our purposes, 
we used Prompt-4 and Prompt-5 with Johnson's $V$ and $I$ filters and 
1024 $\times$ 1024 cameras, giving $604 \times 604$
arcseconds field of view (0.59 arcsec/pix). 
Our PROMPT observations of ASAS-09 took place in January 2010. 
We were able to obtain nearly full $V$ and $I$ light curves in
two weeks. We consider this time
scale as the limit for the time-consistency of the light curve.

ASAS-21 was observed at SAAO in May 2008 with the same instrumentation 
as ASAS-09. The partial light curve again shows a spot-originated 
out-of-eclipse variation. PROMPT observations carried out in June 2009 
were initially performed in order to fill the gaps in the system's 
light curves but it immediately became clear that the new data show 
different light variations and that a consistent model cannot be 
constructed. In July 2009 we used PROMPT to obtain a new light curve, 
which covers all orbital phases. Owing to rapid spot evolution we 
could not use the June and July data together, since again we noticed 
substantial differences in the out-of-eclipse pattern (mainly near 
the secondary eclipse).

The last part of the data comes from SAAO observations taken in 
October 2009 and again shows a different light modulation. Because of 
challenging conditions, the moments of the minima were recorded only in 
the $V$ band, we failed in our attempts to create a separate solution 
based on this data set alone.

The absolute calibration procedure for SAAO data, described in our
paper~2, was based on the catalogue information of 
stars in the observed fields. For PROMPT we obtained additional 
observations of Landolt standard fields 53 ($\alpha$ = 06:52:12, 
$\delta$ = -00:18:00) and 54
\citep[$\alpha$ = 06:52:31, $\delta$ = -00:16:30;][]{lan92}
with both telescopes (Prompt-4 and 5). 
The instrumental magnitudes of the standard stars were corrected for the 
atmospheric extinction and transformed in a traditional manner to the
Johnson's system. In order to improve the calibration for SAAO we compared 
instrumental magnitudes of the stars common in SAAO and PROMPT fields 
of ASAS-09, ASAS-21 and three other systems observed by us with both
instruments. Our final transformations result in the 
magnitudes' absolute uncertainties
$\sigma_V = 0.013$ and $\sigma_I = 0.016$ mag in the
$V$ and $I$ bands respectively. The total number of the data points
collected in every data set for both binaries is shown in Table
\ref{tab_data_no}.

\section{Analysis}

\subsection{Spectroscopic orbits}

\begin{table}
\caption{Barycentric radial velocities of ASAS-09 and ASAS-21}
\label{tab_rv}
\centering
\begin{tabular}{l r r r r r r}
\hline \hline
Date (TDB) & $v_1$ & $\pm$ & {\it O-C} & $v_2$& $\pm$ & {\it O-C} \\ 
-2450000.0 & km~s$^{-1}$ &  &  & km~s$^{-1}$ &  & \\
\hline
{\bf ASAS-09} &  &  &  &  &  & \\
\multicolumn{7}{l}{\it From AAT/UCLES}\\
4837.1911 &  136.75 & 1.75 &  0.34 &  -93.08 & 4.99 & -1.14 \\
4837.2062 &  131.96 & 1.69 &  1.86 &  -87.62 & 4.52 & -1.99 \\
4840.1398 &  -57.14 & 1.95 & -1.39 &  106.51 & 3.94 &  6.17 \\
4840.2420 & -100.56 & 2.21 &  1.17 &  141.33 & 4.75 & -5.03 \\
\hline
{\bf ASAS-21} &  &  &  &  &  & \\
\multicolumn{7}{l}{\it From Radcliffe/GIRAFFE}\\
3903.6543 & -125.31 & 1.33 & 0.37 &  112.00 & 3.20 &-0.53 \\  
3903.6949 & -141.05 & 1.07 & 0.34 &  130.09 & 3.05 &-1.08 \\  
3904.6314 &   32.74 & 2.48 & 2.12 &  -72.70 & 2.93 & 1.42 \\  
3905.5318 &   87.00 & 1.52 &-1.20 & -144.90 & 3.86 &-3.72 \\  
3906.5668 & -135.15 & 1.28 &-1.30 &  123.63 & 2.35 & 4.69 \\  
3906.6109 & -105.44 & 1.40 & 0.57 &   85.47 & 2.94 &-0.03 \\  
3906.6489 &  -67.42 & 1.72 & 2.75 &   51.39 & 3.11 &-4.11 \\  
\multicolumn{7}{l}{\it From AAT/UCLES}\\
4727.0620 &  -97.72 & 1.61 &-2.56 &   76.92 & 3.40 & 0.58 \\
4728.0651 &   95.93 & 3.99 & 0.59 & -148.10 & 2.05 & 1.15 \\
\hline
\end{tabular}
\end{table}

\begin{table}
\caption{Spectroscopic orbital parameters of ASAS-09 and ASAS-21}
\label{tab_orb}
\centering
\renewcommand{\footnoterule}{}
\begin{tabular}{l r l r l}
\hline \hline
&\multicolumn{2}{c}{\bf ASAS-09}&\multicolumn{2}{c}{\bf ASAS-21}\\
Parameter & Value & $\pm$ & Value & $\pm$ \\
\hline
$P$ [d] 		& 0.8974204 & 2.0e-6 	& 0.7024303  & 3.0e-7 \\
$T_0$ [JD-2450000]	& 5205.2949 & 3.0e-4 	& 5009.86436 & 9.0e-5 \\
$K_1$ [km~s$^{-1}$] 	& 126.9 & 1.3		& 126.4 & 1.0 \\
$K_2$ [km~s$^{-1}$]  	& 127.4 & 2.7	 	& 149.7 & 1.4 \\
$\gamma_1$ [km~s$^{-1}$]&  22.7 & 3.1		& -16.6 & 0.6 \\
$\gamma_2-\gamma_1$ [km~s$^{-1}$]& -0.3 & 2.5	&   1.7 & 2.8 \\
$q$ 		 	& 0.996 & 0.023 	& 0.844 & 0.011 \\
$a \sin{i}$ [R$_\odot]$ & 4.510 & 0.052 	& 3.834 & 0.023 \\
$e$ 		 	& 0.0 & (fix) 		& 0.011 & 0.005 \\
$\omega [^\circ]$ 	& --- &  ---  		& 80.   & 30. \\
$M_1 \sin^3{i}$ [M$_\odot]$ & 0.767 & 0.033 & 0.830 & 0.016 \\
$M_2 \sin^3{i}$ [M$_\odot]$ & 0.764 & 0.021 & 0.701 & 0.013 \\
\hline
\end{tabular}
\flushleft
\scriptsize
Note: Values of $P$, $T_0$, $e$ and $\omega$ for both systems are 
taken from the joined spectroscopic and photometric analysis 
performed with PHOEBE. $(\gamma_2-\gamma_1)$ was held fixed to 0
in the further analysis of both systems.
\end{table}

In the analysis of ASAS-09 and ASAS-21 we slightly modified our 
approach as presented in paper~1 and paper~2. We firstly took the 
ASAS light curve (nine year time span) and derived the orbital 
period $P$ and $T_0$ with the fast and stable code 
JKTEBOP \citep{sou04a,sou04b}, based on EBOP 
\citep[\emph{Eclipsing Binary Orbit Program};][]{pop81,etz81}.
This step was especially necessary for the ASAS-09 system.
We also inspected the ASAS light curves for hints of orbital 
eccentricity, but in both cases $e$ was indistinguishable 
from 0 from the ASAS data only.

We used the values of $P$ and $T_0$ to obtain a preliminary 
orbital solution using RV measurements independently of the
light curve. We used a simple procedure, which fits a double-Keplerian 
orbit and minimizes the $\chi^2$ function with a Levenberg-Marquardt 
algorithm. For ASAS-09 we kept the ephemeris and eccentricity fixed 
and fitted for the RV semi-amplitudes $K_1$ and $K_2$ and systemic 
velocities for the primary and secondary $\gamma_1$ and $\gamma_2$. 
For both systems we found $\gamma_1 - \gamma_2$
indistinguishable from 0, so we kept it fixed to 0 in the further analysis.
For ASAS-21 we also set $T_0$, $e$ and the periastron longitude $\omega$ 
free and fitted for them with $K_{1,2}$ and $\gamma_{1,2}$. We found
$e$ to have a small but significant value of $0.03\pm0.02$. It was
later corrected to $0.011\pm0.005$ with the values and uncertainties 
of $P$, $T_0$ and $\omega$ based on the joined spectroscopic
and photometric analysis performed with PHOEBE. After correcting
the values of $e$, $P$, $T_0$, and $\omega$, we ran our code again 
(keeping them fixed) to obtain final values of all spectroscopic parameters. 

The formal RV measurement errors were computed from the
bootstrap analysis of TODCOR maps created by adding randomly 
selected single-order TODCOR maps. For ASAS-09 the formal 
errors are underestimated. In order to obtain a reduced $\chi^2$ of the 
best fit equal to $\sim$1, we added in quadrature an additional error,
which was 0.85 and 3.70 km~s$^{-1}$ for the primary and secondary 
respectively. For ASAS-21 we reached a reduced $\chi^2$ of the best 
fit close to 1, so no additional error was added to the ASAS-21 RV 
measurements. To compensate for the differences in the two 
spectrographs we shifted our RVs obtained with the AAT/UCLES for the 
this system by a constant value of $-4.16\pm2.90$ km~s$^{-1}$. 
This value probably originates in the intrinsic properties of 
the instruments and its uncertainty was included in the further analysis. 

The velocities, their final errors, and $O-C$s from the best-fitting
model are collected in Table \ref{tab_rv}. The errors at the level of 1-5 
km~s$^{-1}$ are probably due to the low SNR, the
rotational broadening, which reaches 44 km~s$^{-1}$ for ASAS-09 
and 61 km~s$^{-1}$ for ASAS-21 (both systems seem to be tidally locked), 
template spectra mismatch,
the activity and, of course, the systematic term (3.70 km~s$^{-1}$
for ASAS-09 secondary) probably originating in the activity itself.
Still, the TODCOR maxima were always well separated, hence 
blending of the spectral lines of the components did not affect 
the velocities. 

In order to verify that the formal least-squares errors of the best
fit parameters are correct, we carried out a bootstrap analysis. 
To this end we created 1000 artificial RV data sets by drawing
replacement RVs from the original RV data sets and fitting the model 
for each such set. These best-fit parameters were then used
to create the distribution for each model parameter and to compute
the corresponding uncertainty. The bootstrap uncertainties were fully consistent 
with the ones from the formal least-squares fitting. 
This most likely means that the formal uncertainties are 
conservative even though the number of single RV datapoints -- 8 for 
ASAS-09 and 18 for ASAS-21 -- is relatively low. Additionally, we
performed a Monte-Carlo simulation to estimate the systematic errors of
derived parameters. The values of fixed parameters were randomly
perturbed by values from a Gaussian distribution with $\sigma$
equal to the given parameter error. The resulting systematic terms,
even if they are at an order of magnitude smaller than the formal ones,
were added in quadrature to the errors from the least-squares fit. 
We believe that this procedure ensures that the error values we 
obtained are well estimated, at least when not
including the influence of the spots (see Section 5.4).

We present the derived spectroscopic parameters for both binaries in 
Table \ref{tab_orb} and our best fitting models in Figure \ref{fig_rv}.
One should note that the RV semi-amplitudes of ASAS-09 differ
by only 0.5 km~s$^{-1}$, which is well within their errors. Thus we deal
with a system composed of (nearly) twin stars. At present only 
three low-mass systems show this situation: \obj{YY~Gem} 
\citep{tor02}, \obj{BD~-22~5866} \citep{shk08} and \obj{LP~133-373} 
\citep{vac07}, with the last one having the largest uncertainty in 
the determination of masses. Two other low-mass binaries -- \obj{GU~Boo} 
\citep{win10} and \obj{Tres-Her0-07621} \citep{cre05} -- have their $q$ 
consistent with 1 within 2$\sigma$. In the sample of 95 binary systems
with their component masses and radii determined with accuracy better 
than 3\%, presented by \citet{tor10}, only 13 systems meet this criterium
(including YY~Gem).

ASAS-21 is on the opposite side of the mass-ratio distribution. 
According to \citet{luc06}, there is a statistically significant narrow 
peak in the $q$ distribution function for $q \ga 0.95$ (the strong
twin hypothesis). \citet{hal03} also refer to a broad peak for 
$q \ga 0.85$, however this is still not confirmed. In the sample presented
by \citet{tor10}, 42\% of systems have their $q \geq 0.95$ within errorbars,
and 76\% have $q \geq 0.85$. ASAS-21 is clearly far from the narrow peak
and only marginally in the wide one. This makes it especially interesting
for testing evolutionary models, because within the same absolute
errorbars it is more difficult and constraining to fit a single 
isochrone to a low-$q$ system than to a $q\sim 1$ system.

\subsection{Spotted light curves}
In order to obtain full and reliable models of our eclipsing binaries, we used 
time-consistent photometric data sets. For ASAS-09 we used only the
PROMPT observations from January 2010 and created a single ''base'' 
model. For ASAS-21 we used the PROMPT data from July 2009 and created a 
family of models. We combined our RVs and $V$ and $I$ light curves
with the \emph{PHysics Of Eclipsing BinariEs} \citep[PHOEBE;][]{prs05} 
code, which is an implementation of the Wilson-Devinney (WD) code 
\citep[with updates]{wil71}. It uses the computed gravitational 
potential of each star to calculate the surface gravity and effective 
temperatures and allows for adding several single spots on both 
components and fit for their size, position, and temperature. 
We used the square-root limb-darkening law \citep{dia92} 
with the coefficients automatically interpolated by PHOEBE from the van 
Hamme LD tables \citep{vHa93}. The reflection albedos for both 
components were held fixed at the value of 0.5 as is appropriate for 
convective envelopes. The gravity brightening coefficient was set to 0.32 
-- the classical value obtained by \citet{luc67}.

\subsubsection{ASAS-09 -- single ''base'' solution}
While creating the ''base'' model of the ASAS-09 binary with PHOEBE, we 
followed our approach presented in paper~2. We firstly improved the
ephemeris, then fitted for the temperatures, inclination, and potentials. 
Later we added a single, large cold spot on the secondary component 
(the one that is eclipsed at $\phi=0.5$) and ''manually'' fitted for its position,
size, and temperature until the out-of-eclipse modulation was reproduced
in a satisfactory way. Finally, we set free the spot parameters, the 
component temperatures and potentials, and the inclination angle.
The initial solutions were obtained separately for PROMPT $V$ and $I$ curves 
combined with RVs. Because we found those initial results consistent, 
we used both the PROMPT light curves and RVs to create the 
final solution and reduce the parameter uncertainties.

The initial values of the temperatures were taken from the empirical 
color-temperature calibration by \citet{cas06} and both component
colors were taken from PHOEBE, which enables one to compute 
separately the flux for every component in every band and also seems to 
keep the $V-I$ color at a stable level, independently from the given 
temperatures. In this way the final values of the temperatures should be 
consistent with those from the models and calibrations. Their
uncertainties also include the absolute photometric calibration errors.

When deciding about the spot configuration, we wanted to keep the model 
as simple and robust as possible. The observed sine-wave modulation can be 
explained by a presence of only one large spot, which reaches across or 
covers the polar regions of the star. The choice of  the component 
for the spot location was determined by the depth of the eclipses and 
the mass ratio. We found that if the spot was on the primary and thus 
was eclipsed at the phase $\phi = 0$ (close to the maximum of the spot's 
appearance), the effective temperature ratio of both components would 
then have to be substantially different from 1 in order to reproduce the 
observed eclipse depths. By placing the spot on the secondary, we keep the 
temperature ratio close to 1, which is expected for stars of nearly 
identical masses.

Before we established that a solution is the ''base'' one, we 
checked if it reproduced other light curves only by changing parameters of 
the spots. We also had to fit for the luminosity scaling factors, in order to account 
for small magnitude uncertainties which originate from the $rms$ of the
absolute photometric calibration. Using PHOEBE, we could force the luminosity 
levels to be decoupled from the temperatures, which we held fixed. In this 
way we were able to reproduce other photometric data sets without changing
any of the stellar parameters. 

The parameters derived from the solution we adopted for ASAS-09 are 
presented in Table \ref{tab_par_09}. From the PHOEBE solution we took the
$\Omega$-potentials, temperatures and inclination, and used them 
with the orbital parameters to derive the absolute masses, radii 
(thus $\log{g}$), luminosities (thus magnitudes), and the distance.
In order to calculate them we used the procedure JKTABSDIM, which is 
available with JKTEBOP. The distance was calculated on the basis
of derived $M_{bol}$, the observed magnitudes in several bands -- ours
$V$ and $I$, and $J$, $H$, $K_S$ from 2MASS \citep{cut03} -- and
various bolometric corrections \citep{bes98,flo96,gir02}. The distance
given in Table \ref{tab_par_09} is a weighted average of eight separately 
calculated results. We reached the best consistency between the results
from various bands for $E(B-V)=0$. As expected, our model predicts
nearly equal values of all the parameters for both components.

\begin{figure*}
\centering
\includegraphics[width=0.89\textwidth]{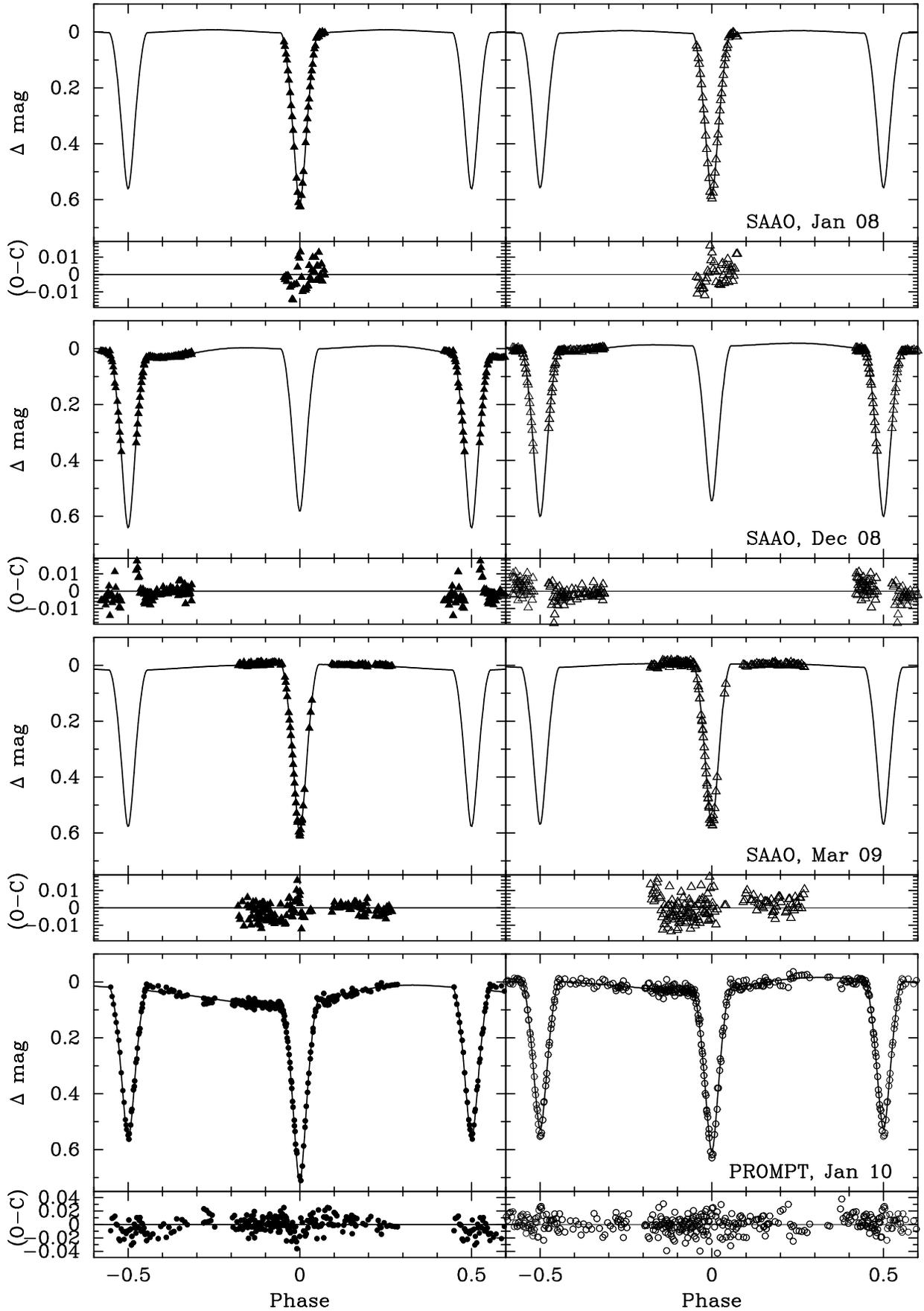}
\caption{The ASAS-09 light curves in the $V$ (left, solid symbols) and $I$ 
	(right, open symbols) bands with their best fitting models. 
	The SAAO data are depicted with triangles and the PROMPT data with circles. 
	The dates of observations are given. Consecutive models differ only by 
	the configuration of spots. The solution from January 2010 is the base 
        one.}
         \label{fig_lc_09}
\end{figure*}

The final models of the ASAS-09 system at various epochs are shown in 
Figure \ref{fig_lc_09}. We present our observations and their respective
models in the chronological sequence. For the January 2008 data, we did not 
apply any spots. The December 2008 solution involves a cold spot on the primary
(not eclipsed) component. We reproduce the March 2009 data set by including
in the model a bright spot (hotter than the surroundings) on the primary 
(eclipsed) component. In this way we can explain that the eclipse 
recorded in March 09 is deeper than that of January 2008. Bright spots
were already used as an explanation for observed out-of-eclipse variations,
for example in \obj{GU~Boo} \citep{lop05}. As we mentioned, the January 
2010 data are reproduced by a model with one large, cold spot, which reaches 
the polar region of the secondary.

One should keep in mind that usually the spot temperature factor 
$T_{spot}/T_{star}$ is degenerate with the spot's angular radius. This is
also the case for ASAS-09. Also, none of the SAAO data sets for ASAS-09
covers both eclipses. Thus the synthetic curves plotted in Fig. \ref{fig_lc_09}
for the SAAO panels should be treated with caution. They were constructed 
only to present how the spot pattern has been changing in time and 
that the observations are reproduced with the ''base'' model from
the January 2010 data.

\begin{figure*}
\centering
\includegraphics[width=0.89\textwidth]{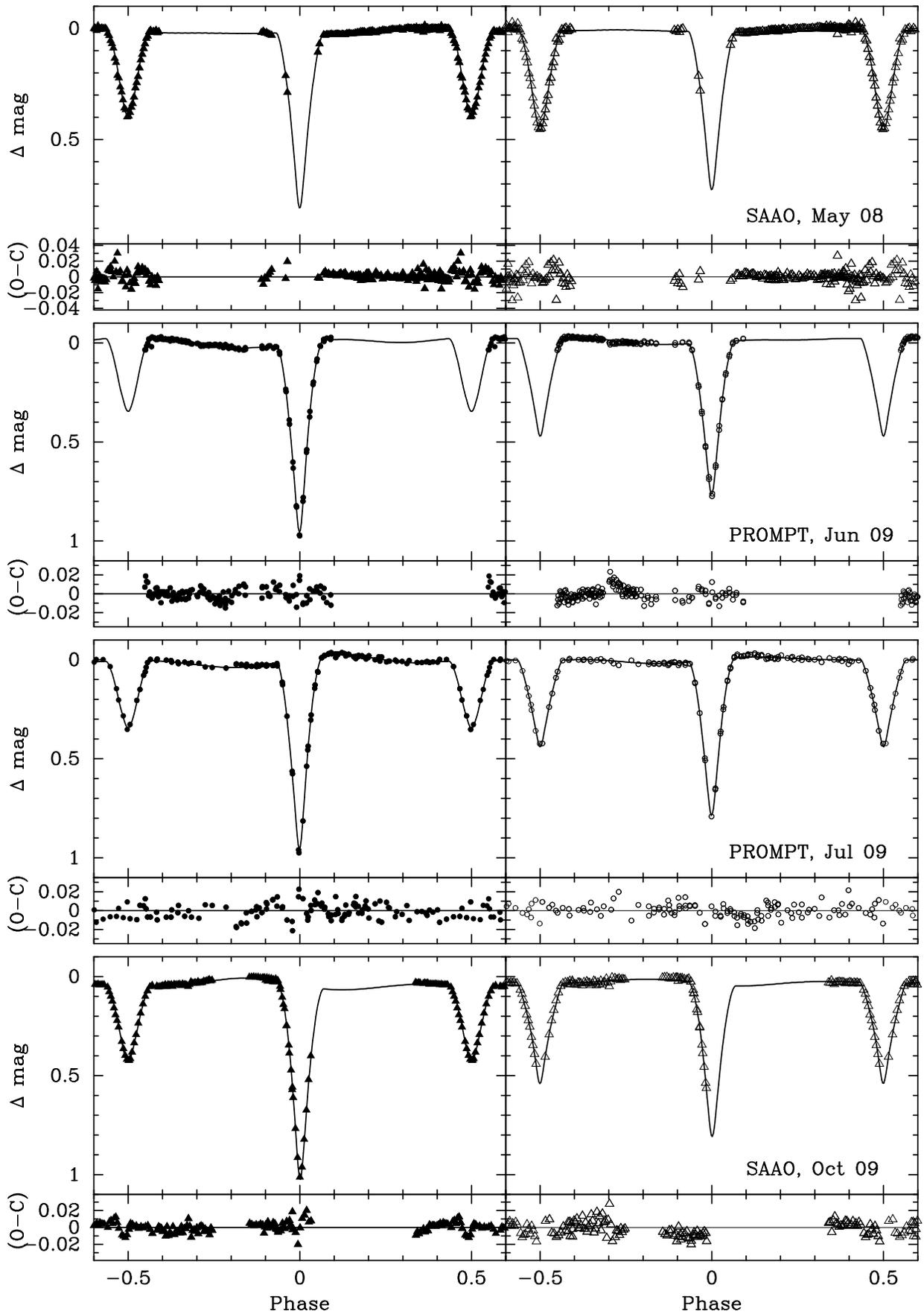}
\caption{The same as Fig. \ref{fig_lc_09} but for ASAS-21. 
	In this case the data from July were selected to derive
	a family of base solutions. Here we present an example with 
        two bright, eclipsed spots.}
         \label{fig_lc_21}
\end{figure*}

\begin{table}
\centering
\caption{Absolute physical parameters adopted for ASAS-09\label{tab_par_09}}
\begin{tabular}{l r l r l}
\hline \hline
{\bf ASAS-09} & \multicolumn{2}{c}{Primary} & \multicolumn{2}{c}{Secondary}\\
Parameter & Value & $\pm$ & Value & $\pm$ \\
\hline
inclination $[^\circ]$ &\multicolumn{4}{c}{86.55 $\pm$ 0.20}\\
$a$ [R$_\odot]$  	&\multicolumn{4}{c}{4.519 $\pm$ 0.054}\\
Mass [M$_\odot]$ 	& 0.771 & 0.033 & 0.768 & 0.021 \\
$\Omega$-potential	& 6.866 & 0.040 & 6.887 & 0.040 \\
Radius [R$_\odot]$ 	& 0.772 & 0.012 & 0.769 & 0.013 \\
$\log{g}$ 		& 4.550 & 0.013 & 4.554 & 0.011 \\
$v_{synch}$ [km s$^{-1}$]&  43.5  & 0.7 & 43.2  & 0.7 \\
$V-I$ [mag] 		& 1.170 & 0.13   & 1.172 & 0.13 \\
$T_{eff}$ [K]		&  4360 &  150   & 4360  & 150 \\
$M_{bol}$ [mag] 	&  6.53  &  0.10  & 6.54   & 0.10\\
$M_V$ [mag] 		&  7.22  &  0.16  & 7.23   & 0.16 \\
distance [pc]		&\multicolumn{4}{c}{151 $\pm$ 9}\\
\hline
\end{tabular}
\end{table}

\begin{table}
\centering
\caption{Potentials and inclination angle for the six solutions 
for ASAS-21 each with a different spot configuration. The last line is for
the weighted averages and their uncertainties adopted for the base solution
of ASAS-21.}\label{tab_sixmodels}
\begin{tabular}{l l r l r l r l}
\hline \hline
No.&Config.& $\Omega_1$ & $\pm$ & $\Omega_2$ & $\pm$ & $i$ [$^\circ$] & $\pm$ \\ 
\hline
1 & 2bn-1bn & 5.365 & 0.039 & 5.496 & 0.027 & 88.66 & 0.7 \\
2 & 1be-1bn & 5.290 & 0.051 & 5.587 & 0.058 & 87.43 & 0.4 \\
3 & 1be-2be & 5.434 & 0.041 & 5.705 & 0.048 & 87.29 & 0.3 \\
4 & 2bn-2be & 5.409 & 0.037 & 5.533 & 0.045 & 87.39 & 0.7 \\
5 & 1be-1c  & 5.333 & 0.067 & 5.617 & 0.070 & 88.17 & 0.8 \\
6 & 1be-2c  & 5.316 & 0.041 & 5.707 & 0.051 & 86.99 & 0.3 \\
&&&&&&&\\
\multicolumn{2}{c}{\it Adopted} & 5.368 & 0.078 & 5.572 & 0.135 & 87.35 & 1.31 \\
\hline
\end{tabular}
\flushleft
\scriptsize
Note: The name of the configuration encodes the character and localization of the spots.
The first part (3 characters) describes the spot responsible for the brightening around 
$\phi\sim 0.1$ and the second (2 or 3 characters) -- for the maximum around $\phi\sim 0.5$. Digit
"1" means the spot is on the primary, and "2" -- on the secondary; "b" stands for
bright spots, "c" for cold; "e" stands for an eclipsed spot, "n" for non-eclipsed
(visible during an eclipse).
\end{table}

\begin{table}
\centering
\caption{Absolute physical parameters adopted for ASAS-21\label{tab_par_21}}
\begin{tabular}{l r l r l}
\hline \hline
{\bf ASAS-21} & \multicolumn{2}{c}{Primary} & \multicolumn{2}{c}{Secondary}\\
Parameter & Value & $\pm$ & Value & $\pm$ \\
\hline
inclination $[^\circ]$  &\multicolumn{4}{c}{ 87.4 $\pm$ 1.3 }\\
$a$ [R$_\odot]$  	&\multicolumn{4}{c}{ 3.838 $\pm$ 0.023}\\
Mass [M$_\odot]$ 	& 0.833 & 0.017 & 0.703 & 0.013 \\
$\Omega$-potential	& 5.368 & 0.078 & 5.572 & 0.135 \\
Radius [R$_\odot]$ 	& 0.845 & 0.012 & 0.718 & 0.017 \\
$\log{g}$ 		& 4.506 & 0.012 & 4.574 & 0.020 \\
$v_{synch}$ [km s$^{-1}$]& 60.8 &  0.5  &  51.7 & 0.9 \\
$V-I$ [mag] 		&  1.15  &  0.21  &  1.39  & 0.30 \\
$T_{eff}$ [K]		&  4750  &  150   &  4220  & 180 \\
$M_{bol}$ [mag] 	&  5.96  &  0.15  &  6.85  & 0.19\\
$M_V$ [mag] 		&  6.35  &  0.14  &  7.65  & 0.21 \\
distance [pc]		&\multicolumn{4}{c}{126 $\pm$ 8}\\
\hline
\end{tabular}
\end{table}

\subsubsection{ASAS-21 -- a family of solutions}

The case of ASAS-21 was more complicated than ASAS-09. Firstly 
because of a measurable eccentricity, which manifests itself in 
a separation of eclipses different than 0.5$P$.
In these cases the zero-phase moment $T_0$ is defined in PHOEBE
in a specific way \citep[for details see paper~1 or][]{prs05} and in
general does not coincide with the primary eclipse. For ASAS-21
the eclipses occur at phases 0.99968 (or -0.00032) for the primary and
0.50032 for the secondary eclipse. This deviation is not visible in
Fig. \ref{fig_lc_21}, but is measurable in PHOEBE. It is however surprising 
that non-zero eccentricity exists in such a system.
Secondly, the out-of-eclipse patterns for the ''base'' data set were
more complicated and did not allow us using a single spot. 
We decided to construct a ''base'' model using the July 2009 PROMPT 
data, because in July we obtained the full phase coverage. 

For the July set we excluded the scenario of two cold spots, 
whose appearance is responsible for the brightness
minima around $\phi$=0.3 and -0.2. On tidally locked components, the 
brightness minimum caused by the spot would last longer than half the 
orbital period. This is not the case for the minimum around $\phi$=0.3,
which seems to last for about $0.4 P$. The component on which such a spot 
could exist would have to rotate faster than in the case of tidal locking 
and would not explain why the same brightness modulation is observed
in every consecutive cycle. Thus we conclude that in the July data 
either one or both brightenings (around $\phi=0$ and $0.5$) are probably
caused by a bright spot. The model of a spot in PHOEBE/WD is
however somewhat simplistic, so we cannot exclude a more complicated 
pattern of cold, dark spots.

There are two possibilities, one that in July 2009 we observed a modulation 
coming from two bright spots and the other that a modulation is caused by  
one bright spot superimposed on sine-like variations caused by a large cold spot 
similar to the one from the January 2010 data of ASAS-09. We concluded that the 
maximum appearance of this large cold spot should occur close to phase 0 (thus 
brightening near 0.5) not -0.4 (brightening around 0.1), since the former 
configuration usually reproduced the observations better than the latter.
We produced a series of solutions by putting various spots on various 
components. We succeeded to derive a satisfactory model for every 
configuration with two bright spots (four cases) and with a bright eclipsed 
spot on the primary component when accompanied by a cold one
(on either component; two cases).

Every single model was prepared in the same way as the ''base'' model 
of ASAS-09. In order to arrive at the ''base'' solution of ASAS-21, 
we took weighted average values of the $\Omega$-potentials, inclinations 
and temperatures. We used the formal errors as weights while computing 
the averages. As the uncertainties of the average value we adopted the 
maximum difference between the average value and the set possible solutions. 
The exception were the temperatures, for which the photometric calibration 
is the main source of uncertainty (the typical formal error taken from 
PHOEBE was at the level of 10~-~20~K). We believe that this approach 
constitutes a conservative way of accounting for systematic 
errors in the parameters due to an uncertain spot configuration. 
All single values of the potentials and inclination for all six allowable 
solutions are listed in Table \ref{tab_sixmodels} with the values
adopted for the ''base'' model. The resulting values of parameters of 
the ''base'' model and their uncertainties were incorporated into 
JKTABSDIM and the absolute physical parameters were calculated. They are 
listed in Table \ref{tab_par_21}. The errors of temperatures, 
$M_{bol}$, $M_V$ and $V-I$, as for ASAS-09, also include the uncertainty
of the absolute photometric calibration. As for ASAS-09, the distance 
was calculated from several bolometric corrections, but for ASAS-21 
we reached the best consistency for $E(B-V) \sim 0.1 - 0.15$. 

In Figure \ref{fig_lc_21} we show all data sets in a chronological 
sequence. The solution for May 2008 was acheived by adding two cold 
spots to the ''base'' model and by fitting for their parameters. 
The data from June 2009 are represented by the ''base'' model with 
two bright eclipsed spots, which is also plotted 
in the July panel. These models differ from each other by the temperature 
factors of spots. It supports the type of the July 2009 solutions with two bright spots, 
since the observed differences in the brightenings levels would be hard to explain 
with the evolution of a large cold spot responsible either for the
brightening near $\phi=0.1$ or especially, 0.5. However, we did not exclude the 
latter in our aforementioned calculations in order not to underestimate
the uncertainties. We also estimate that the typical time scale of spot 
evolution in the ASAS-21 system is about three weeks. For the remaining
data set -- October 2009 -- we also show a model with two bright spots. A spot on the 
primary is responsible for the brightening around $\phi=-0.2$, thus it is
eclipsed. It is possible that the October pattern emerged from the July 
one and the spots moved in longitude, but the possibility of changing the 
location on a tidally locked component is disputable.
Again, in cases where a certain eclipse is not covered by the data,
one should treat the synthetic light curve with a bit of caution.

\section{Chromospheric activity}

\subsection{X-ray emission}

The presence of stellar spots on the surface of both systems' 
components bears testimony to their substantial level of magnetic 
activity. This can be also indicated by the X-ray emission, or
the X-ray-to-bolometric luminosity ratio $L_X/L_{bol}$.
Both systems appear in the ROSAT All-Sky Bright Source Catalogue
\citep[1RXS;][]{vog99} with the observed total fluxes of 
$0.396\pm 0.062 \cdot 10^{-12}$ erg cm$^{-2}$ s$^{-1}$ and
$0.714\pm 0.102 \cdot 10^{-12}$ erg cm$^{-2}$ s$^{-1}$ for
ASAS-09 and ASAS-21 respectively, which gives 
$L_X = 1.08\pm0.21\cdot 10^{30}$ erg s$^{-1}$ and
$L_X = 1.36\pm0.23\cdot 10^{30}$ erg s$^{-1}$ for
the distances from Tables \ref{tab_par_09} and \ref{tab_par_21}.

The high level of activity is claimed to be driven by the stellar rotation.
For single stars the magnetic activity has been found to be correlated 
with stellar rotation \citep{fle89,bou90}, but no such correlation has been 
found for binaries \citep{fle89}, which are usually faster rotators. 
However, \citet{del98} have proven that for the field M dwarfs, 
$L_X/L_{bol}$ saturates already at $v_{rot}\sin{i} > 5$ km~s$^{-1}$,
so most of the low-mass DEBs should fall into the saturated X-ray regime
where no correlation between activity and rotation exists.
The general fact is that fast rotators, like tidally locked short-period
binaries, show a higher level of X-ray emission than slowly rotating stars
of the same mass. 

\citet{lop07} has found a clear correlation for low-mass stars 
between $L_X/L_{bol}$ and the observed radii excess defined as 
$\Delta R = (R_{obs}-R_{mod})/R_{mod}$, where $R_{obs}$ and 
$R_{mod}$ are respectively an observed radius and a radius of 
a star of given mass theoretically predicted by the evolutionary models 
of \citet{bar98}. The inflated radii of components of late-type 
binaries are their main characteristic and are observed for many
years \citep[see Fig. 1 in][]{lac77}. Recently, some successful efforts
of an explanation involving the influence of rotationally driven magnetic
activity on the effectiveness of convection and mixing have been made
and seem to have strong support in the recent burst in the number
of discovered and characterized low-mass eclipsing binaries.
This also includes a $P=8.43$ d system \obj{T-Lyr1-17236}, which even
if rotationally synchronized should exhibit relatively low rotational velocity
which therefore should not induce a high level of activity \citep{dev08}.
However, the authors did not find a conclusive evidence for the non-activity
of this system, especially considering the signatures of $H_\alpha$ emission 
and the 0.1 mag level of the photometric data scatter. Most of the spotted
systems -- including ASAS-09 and -21 -- show out-of-eclipse modulation
of this level or lower (see Fig. \ref{fig_spot_lc}).

\begin{table*}
\caption{Observed and theoretically predicted radii of ASAS-09 and
	ASAS-21 components and their $L_X/L_{bol}$ calculated from the
	values from Tables \ref{tab_par_09} and \ref{tab_par_21} and under
	assumption of no, linear and quadratic correlation between $L_X$ and 
	$v_{rot}$. The numbers in parenthesis show the uncertainty.}\label{tab_X_R}
\centering
 \begin{tabular}{lcccccc}
 \hline \hline
Component & $R_{obs}$ & $R_{mod}$ & $\Delta R$ &
\multicolumn{3}{c}{$L_X/L_{bol} \cdot 10^4$}\\
	& [R$_{\odot}$] & [R$_{\odot}$] & [\%]  & $(\nsim v_{rot})$ & $(\propto v_{rot})$  & $(\propto v_{rot}^2)$\\
\hline
ASAS-09 A & 0.772(12) & 0.699(28)  & 10.4(6.4) & 7.27(31) & 7.27(31) & 7.28(30) \\
ASAS-09 B & 0.769(13) & 0.696(18)  & 10.5(4.8) & 7.33(33) & 7.33(33) & 7.32(33) \\
ASAS-21 A & 0.847(12) & 0.751(15)  & 12.8(3.9) & 5.40(28) & 5.79(27) & 6.19(26) \\
ASAS-21 B & 0.718(17) & 0.646(10)  & 11.1(4.5) &12.36(29) &11.45(30) &10.55(32) \\
\hline
 \end{tabular}
\end{table*}

In Table \ref{tab_X_R} we present our estimates of $L_X/L_{bol}$ 
based on our measurements of the radii and bolometric luminosity, radii 
theoretically predicted by $t=1$ Gyr $Z=0.02$ isochrone of \citet{bar98}, 
and three assumptions about the X-ray luminosity and rotational velocity 
considered in \citet{lop07}: none, linear and quadratic correlation. 
In the first case (no correlation) we followed the assumption that the X-ray 
emission is divided equally between components. The values suggest 
that ASAS-21 is the more active of the two systems. We see that 
both systems exhibit radii inflated by about 10--12\%, while their 
X-ray emission rates are substantially different, including the given errors.
This can be explained only if no correlation between $L_X$ and 
$v_{rot}$ exists and the X-ray emission is not equally divided between the
components of one system. It is also very likely that the level of X-ray 
emission, and thus the magnetic activity, varies in time and is not dependent 
on the stellar mass, at least in rapidly rotating, tidally locked binaries. 
This claim seems to be supported by the ASAS-09 and -21 light curves,
with the out-of-eclipse modulation pattern varying in time. However, no
secure conclusions can be made from our data only. An intriguing
question would be whether the radii change with the varying activity level.
In order to answer this question one could combine simultaneous time series 
of X-ray and multi-band photometric observations, preferably 
also with high-precision RV measurements. To the best of our knowledge, 
no such efforts have been carried out. 

\subsection{$H_\alpha$ emission}
One of the most important spectral indicators of chromospheric activity 
is the emission in the Balmer $H_\alpha$ line. Many known low-mass 
stars exhibit such an emission, which is often variable and decreases 
in time for single stars \citep{haw99}. Our low-SNR spectra
make the $H_\alpha$ line identification quite challenging and impossible
in the case of other activity indicating spectral emission lines. $H_\beta$
was in the range of AAT/UCLES only but no conclusive identification was
made. The Ca II H, K, and helium 4027 and 4473 \AA\,lines were out of the 
range of both spectrographs with our setup.

In two of our AAT/UCLES spectra of ASAS-09 we identified the
$H_\alpha$ emission line of both components, but we could make 
reliable measurements only of the one taken at HJD = 2454837.2062. 
We found the equivalent widths to be $0.25 \pm 0.15$
and $0.15 \pm 0.1$ \AA\,for the primary and secondary, 
respectively. The impossibility of obtaining reliable measurements 
in this case does not allow us to draw conclusions about the 
variability of the $H_\alpha$ emission line. However, we cannot 
rule it out.

We found $H_\alpha$ to be in emission for ASAS-21 in several 
Radcliffe/GIRAFFE and AAT/UCLES spectra. In the former case the line 
was close to the edge of an \'echelle order, so no meaningful measurements 
could be done. Emission from the primary was certain but 
undetectable for the secondary. This may mean however that the 
absorption feature of $H_\alpha$ was filled-in to the level of continuum. 
In one of the AAT/UCLES spectra we found an emission from both components
with the equivalent widths of $0.4 \pm 0.1$ and $0.5 \pm 0.1$ \AA\,for 
the primary and secondary respectively. From the comparison with the SAAO 
spectra we may conclude that at least for the secondary the 
$H_\alpha$ emission varies.

\subsection{Stellar spots and their evolution}\label{sec_spots}

Considering the ''base'' model of ASAS-09 we see that probably only one 
component was heavily spotted at the time of observations. From the 
model of the spot, we deduce that the polar region of the
secondary is covered. The existence of these features is not 
straightforward to deduce and understand. The first theoretical predictions 
of these features were done by \citet{sch92} according to whom polar 
spots in rapidly rotating stars are caused by the Coriolis force that
overpowers the buoyancy. They can be used to explain sine-like modulations
because an equatorial spot is not visible at certain phases and therefore
would not modify the ''clean'' light curve. These sine-like modulations and 
polar spots were noted for low-mass CM~Dra \citep{lac77}, LP~133-373 
\citep{vac07}, BD~-22~5866 \citep{shk08}, \obj{NSVS~02502726} 
\citep{cak09}, \obj{ASAS~J082552-1622.8} (Paper 2) or a  
1.13 + 1.05 M$_{\odot}$ binary \obj{UV~Leo}, \citep{mik02,prs05}. 
Further photometric investigations of this system revealed that 
the spot pattern also had changed on a time scale of single weeks 
(compare \citealt{kju07} with \citealt{hec09}). 
It is also noticeable that different 
studies resulted in different values of the ratio of radii \citep[see][]{hec09}.
Another example of short-scale spot variations is probably a 
1.38 + 1.3 M$_{\odot}$ pre-main-sequence binary 
\obj{ASAS~J052821+0338.5} \citep{ste08}, where the authors
report significant changes in the light curve during a 110-day monitoring.
A longer term evolution on time scales of single years is reported for 
detached BD~-22~5866 \citep{shk08} and \obj{RS~CVn} \citep{rod94}, 
and semi-detached \obj{ES~Cnc} \citep{pri08} and \obj{RT~Lac} 
\citep{cak03}. The variability on time scales of several hundreds of days is
also seen in the cases of ASAS-09 and ASAS-21. Their long-cadence ASAS
light curves exhibit semi-periodic variations with $P\sim1000$ and 
1600 d, respectively.

All this suggests that a rapid and long-time spot evolution is as
common for active binaries as it is for single stars. This influences
the derived properties, in particular the radii and inclination, as  
was shown in a recent study of GU~Boo by \citet{win10}.
They analyzed their new multi-band CCD and photoelectric measurements, 
reanalyzed photometric data of \citet{lop05}, and found that the 
varying spot pattern implies a discrepancy of about 2\% 
between the three data sets they used. Also changing the spot configuration 
(moving a spot from the primary to the secondary for instance) 
for the same data set \citep[from][]{lop05} may change the resulting 
radii by over 3\% and the masses by 2\%. 
We hold that this was not the case at least for ASAS-09, because
a different spot configuration would lead to less probable values of 
the temperatures. As for ASAS-21, averaging over several models 
already led to $\sim$2\%  uncertainties. It is an open question whether
the discrepancies in radii, like those found by \citet{win10}
are caused only by the models of spots that were used or are a 
real manifestation of the varying level of activity. The latter 
however should not influence the orbital inclination, 
and following from this, the masses. If the variation in radii is real, 
it should not be at the level higher than about 1-2 \%, which would be 
difficult to detect in the case of binaries described here. 
To observe it, one would require a series of determinations of
the component radii with about 0.5\% precision from the data 
gathered during several observing epochs. As pointed out by 
\citet{win10}, there is a clear need of long-term
monitoring and derivation of the absolute parameters of active
systems in order to reach a better understanding of their nature. 


\subsection{Influence of spots on the precision}

Stellar spots on rapidly rotating stars also influence the measured
radial velocities. Because the spot pattern varies over single weeks, 
the RV and photometric observations should be done nearly 
simultaneously, which is not the case in our study. The RV measurements
of ASAS-09 at the AAT/UCLES were carried out after about a 
month from the SAAO observing run in December 2008, but the December 
2008 light curve is far from complete and provides no strong constrains 
of the configuration of spots. For this reason we did not perform our 
RV fits with the spots (as is possible in PHOEBE), but only 
checked which impact this could possibly have on the RV analysis and 
the derived parameter precision.

\begin{figure*}
\centering
\includegraphics[width=\textwidth]{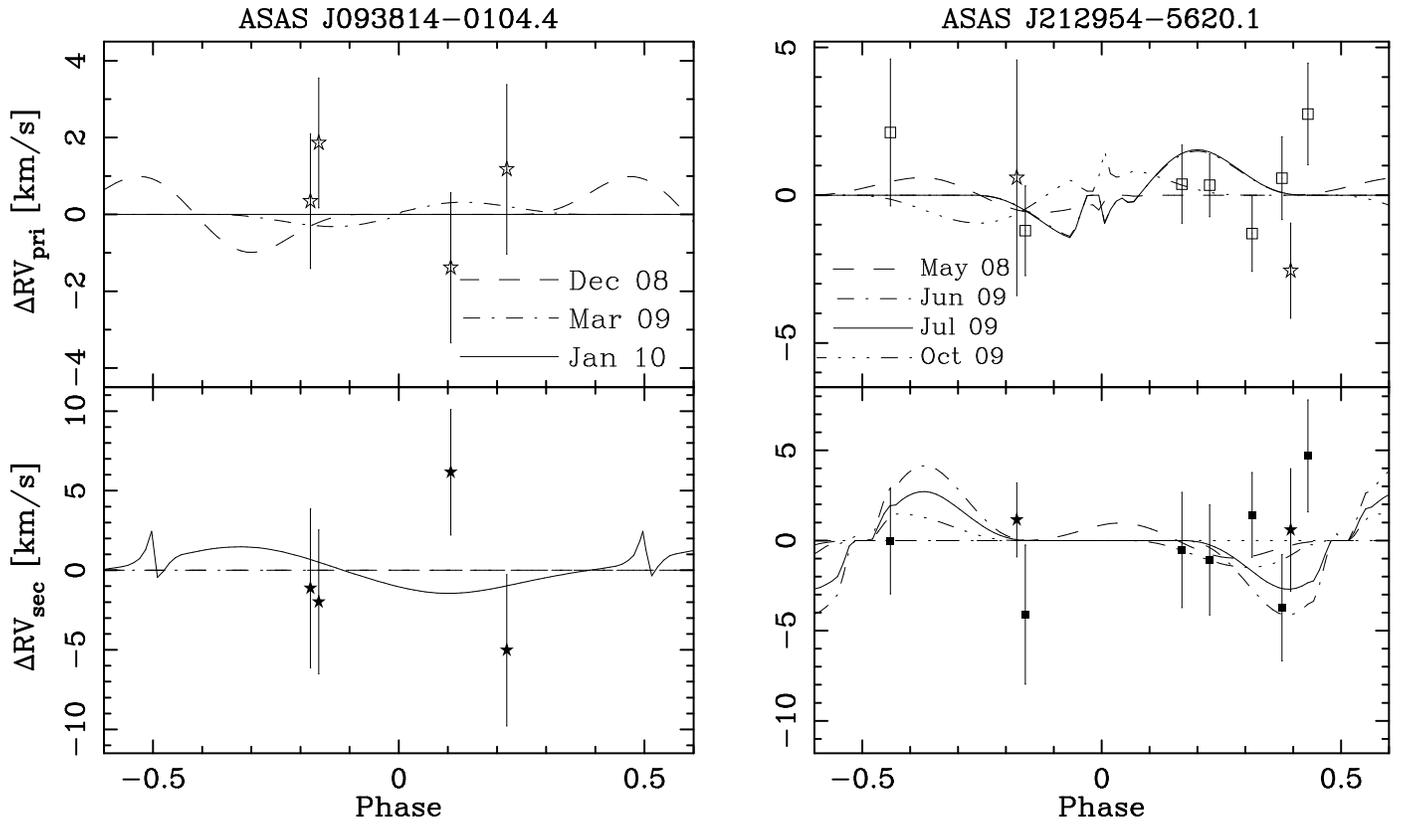}
\caption{Influence of stellar spots included in our model on
the radial velocity curve of the primary (top) and secondary (bottom)
components of ASAS-09 (left) and ASAS-21 (right) systems. 
ASAS-09 models were constructed with one spot and ASAS-21 with two
spots (one on each component). Solid lines depict the base solution
which in the case of ASAS-21 is represented by the configuration, 
with two eclipsed spots. Points with errorbars are the residua of the
orbital fit. Their coding is the same as in Fig. \ref{fig_rv}.
}\label{fig_spot_rv}
\end{figure*}

\begin{figure*}
\centering
\includegraphics[width=\textwidth]{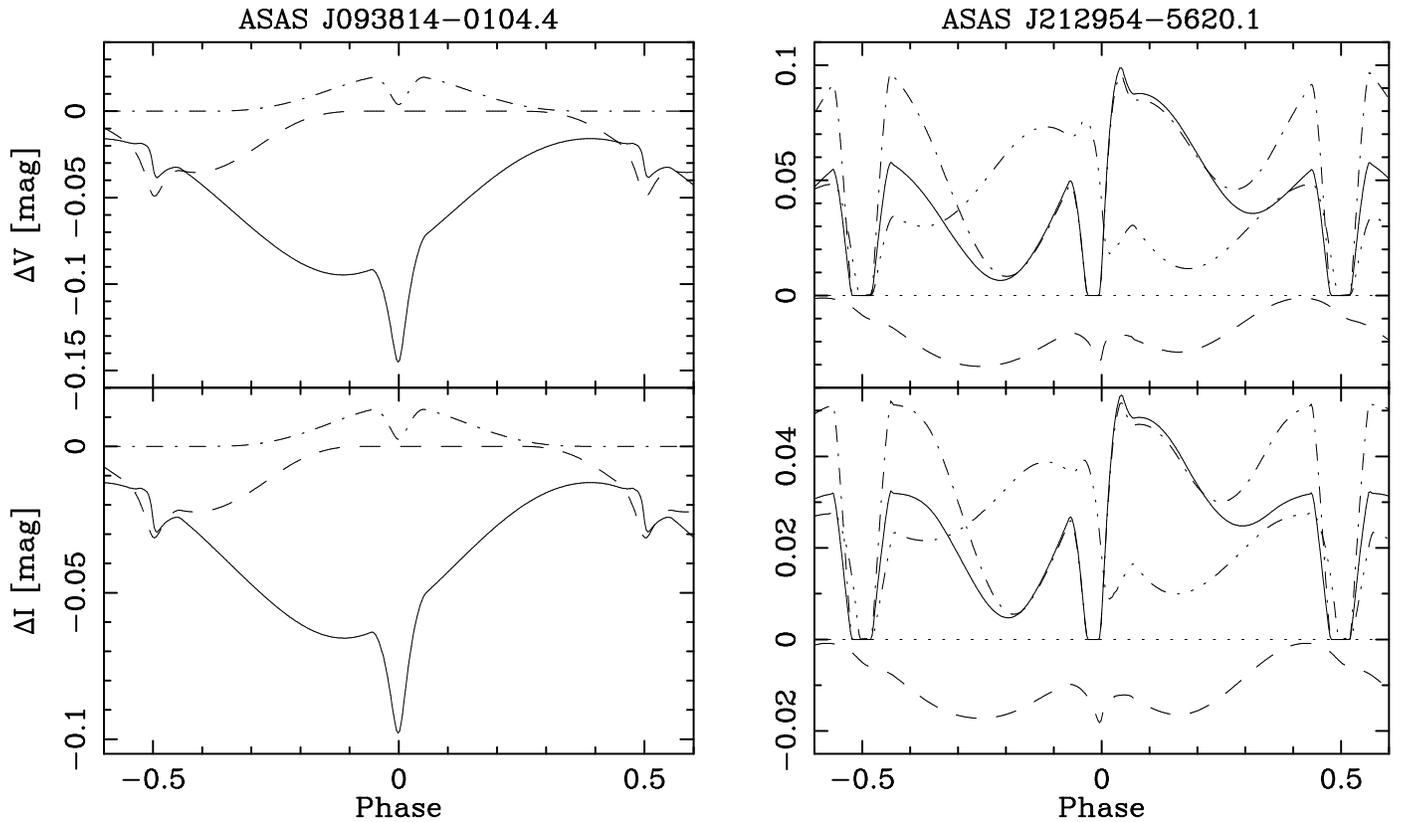}
\caption{Influence of stellar spots included in our model on
the $V$ (top) and $I$ band (bottom) light curves of ASAS-09 
(left) and ASAS-21 (right) systems. The line encoding is the 
same as in Fig \ref{fig_spot_rv}. For the purpose of clarity
the zero-level is shown at ASAS-21 plots only with a dotted line.}
\label{fig_spot_lc}
\end{figure*}

We show the spot-originated RV modulation 
in Figure \ref{fig_spot_rv} and the influence
on the light curves from the models in Figure \ref{fig_spot_lc}. 
The differential RV curves for the primary component 
are shown in the upper panels and for the secondary components in 
the lower ones. The base solutions are depicted with solid lines. 
The $\Delta$RV curves are compared with the orbital fit residuals 
and errorbars. The point coding is the same as in Fig. \ref{fig_rv}.
For ASAS-09 we can see no differences in the RVs either, because our 
models were constructed with only one spot on a certain component.

For ASAS-21 the bright (eclipsed) spot on the primary for the June
solution is very similar to the analogical spot from the July model 
(with two eclipsed spots), so not much of a difference is seen between 
those cases of $\Delta$RV for the primary nor of $\Delta V$ or 
$\Delta I$ around $\phi\sim0.1$ (Fig.\ref{fig_spot_lc}). An evolution
of the spot on the secondary is obvious however and in our opinion the 
scenario of two bright spots in the system is the best one to
explain this behavior in a simple way. Figure \ref{fig_spot_lc}
shows that at longer wavelengths ($I$ band) the influence is
much smaller, thus RVs measured near-IR should not be significantly
affected by the spots \citep[as confirmed for example by][]{fig10}. 
It also means that spectrographs with a long spectral range will produce 
RV measurements with larger errors if the spot information
is excluded from the analysis. It would be presumably possible to
find a correlation between the \'echelle spectrum order numbers
(e.g. their central wavelength) and RV measurements from single 
orders, which would enable us deduce some information about the 
spot itself. It is also worth noting how the presence of a cold 
spot on the secondary component of ASAS-09 in the January 2010 model 
makes the primary eclipse deeper.

From Figure \ref{fig_spot_rv} we can see that the RV 
errorbars and their spread is bigger than the possible influence of spots.
We can conclude therefore that the condition of obtaining $\chi^2$ of the 
orbital fit equal to 1, mainly through adding a systematic term, 
is sufficient to compensate for the influence of stellar activity. 
This was already shown in paper 2. We believe that the uncertainties 
of the derived parameters given in Tables \ref{tab_par_09} and 
\ref{tab_par_21} are well estimated and reliable even though that
the spectroscopic and photometric observations were not carried out
simultaneously and the spot pattern has changed. Likely
obtaining all data at (nearly) the same time would allow for a
better precision of the results.

\section{Age estimae and evolutionary status}

\subsection{Kinematics}
We calculated the galactic velocities $U,V$, and $W$\footnote{Positive 
values of $U$, $V$ and $W$ indicate velocities toward the Galactic center, 
direction of rotation and north pole respectively} with regard to the 
local standard of rest \citep[LSR;][]{joh87}. We took our values of
distance and systemic velocity (with uncertainties, and assuming 
$\gamma_1 - \gamma_2 = 0$), and applied the positions and proper 
motion values from the UCAC3 \citep{zac10}. We obtained 
$U=-4.5 \pm 2.9$, $V = -21.0 \pm 3.0$, and $W = 4.5 \pm 3.1$ 
km~s$^{-1}$ for ASAS-09, which puts it definitely in the young disk,
at the edge of a region occupied in the $(U,V,W)$ space by the Pleiades
moving group \citep[PMG, a.k.a. the Local Association;][]{sea07,zha09}.
The PMG is claimed to be between 20 and 150 Myr old \citep{mon01}
with a metal abundance bellow solar \citep{zha09}.

The values of galactic velocities we obtained for ASAS-21 --
$U=-40.6 \pm 2.6$, $V = -56.1 \pm 4.0$ and 
$W = 4.3 \pm 1.1$ km s$^{-1}$ -- put this system kinematically
in the so called Hercules Stream \citep{sea07,zha09}. This feature 
is a mixture of young and old stars, which show kinematic 
characteristics typical for the thick galactic disk \citep{ben07}. 
If so, the ASAS-21 system can be as old as 8 Gyr 
\citep[lower limit for the thick disk age;][]{jim98} or more.
Considering the kinematics, ASAS-21 appears to be older
than ASAS-09. However, the measured excess in radii and
the $L_X/L_{bol}$ ratio (Table \ref{tab_X_R}) suggest that it is 
more active than ASAS-09. This is possibly connected to the
smaller separation and the fact that the secondary of ASAS-21 is less
massive than the ASAS-09 components, which for single stars 
would directly mean a higher level of activity. A smaller separation
also indicates higher rotational velocities, which in turn may additionally
increase the activity level of both components, if such a correlation
still exists for high $v_{rot}$.
 
\subsection{Isochrones}

\begin{figure*}
\centering
\includegraphics[width=0.85\textwidth]{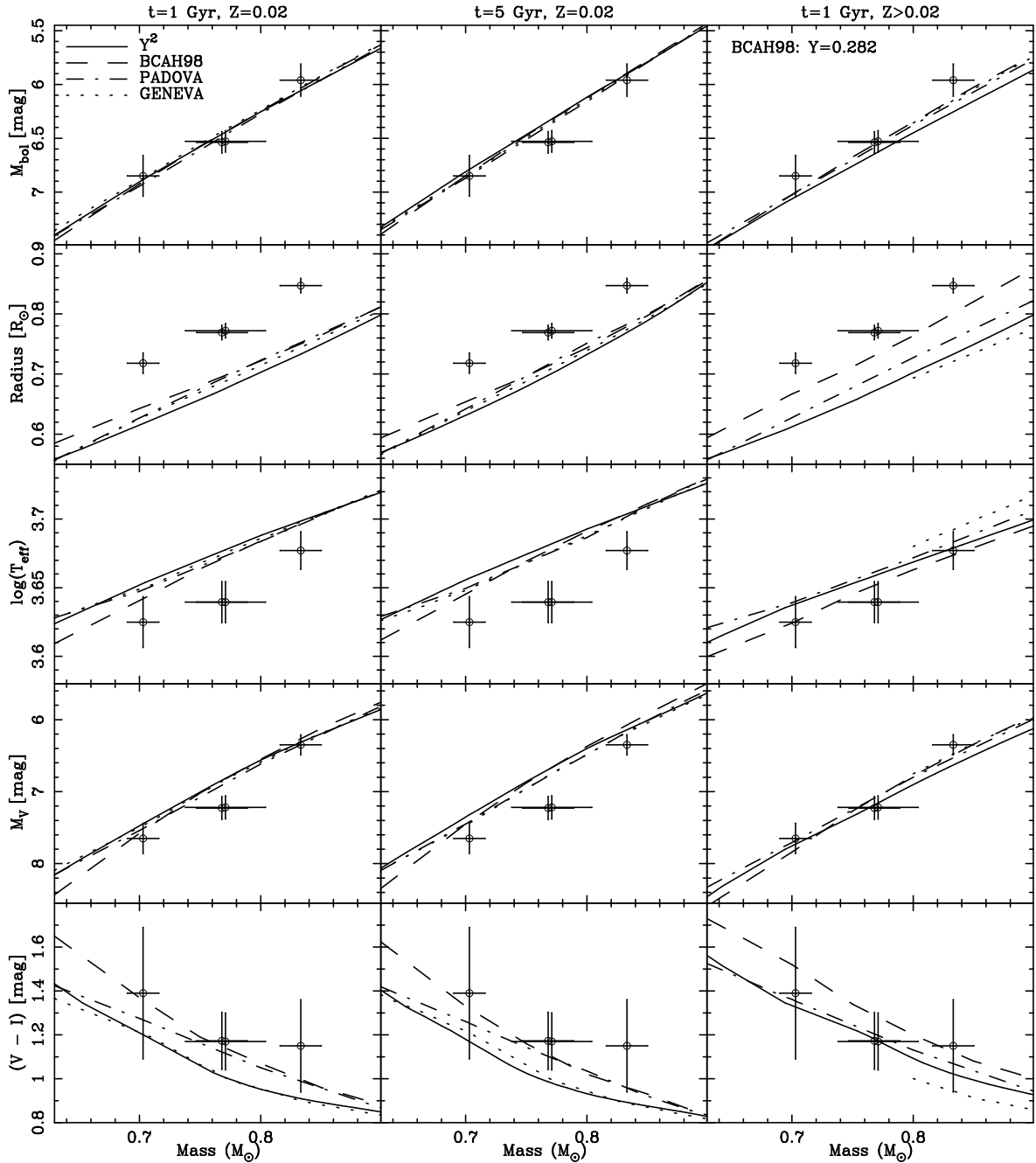}
\caption{Comparison of various parameters as a function of mass 
	for ASAS-09 (inner pair) and ASAS-21 with several sets of isochrones.
	Left column is for the age of 1 Gyr and nearly-solar metallicity, middle 
	column is for 5 Gyr and $Z=0.02$, right column is for 1 Gyr and 
	metal-enriched composition: 0.03 for PADOVA, and 0.04 for Y$^2$ and 
	GENEVA (available for $M \ge 0.8$ M$_\odot$). For the comparison 
	we also show an isochrone of $Y=0.282$ for BCAH98.
	\label{fig_evo_0921}}
\end{figure*}

We compare our results with several popular and widely used sets of
theoretical evolutionary models. For our purposes we chose 
Y$^2$  \citep{yi01,dem04}, BCAH98 \citep{bar98}, PADOVA 
\citep{gir00,mar08} and GENEVA \citep{lej01}. In Figure \ref{fig_evo_0921} 
we show our measurements in the mass $M$ vs. (from top to bottom) 
the bolometric luminosity $M_{bol}$, radius $R$, logarithm of the effective 
temperature $\log{T_{eff}}$, absolute $V$ magnitude $M_V$,
and $V-I$ color. ASAS-09 is the close pair near $M=0.77$ M$_\odot$. 
In order to show the probable properties of the researched systems,
we made our comparison for three cases: (1) for nearly-solar metallicity 
$Z=0.02$ and age $t = 1$ Gyr; (2) for $Z=0.02$ and $t=5$ Gyr;
(3) $t=1$ Gyr and metal abundance above solar, which varies depending
on the set of isochrones: 0.03 for PADOVA, and 0.04 for 
Y$^2$ and GENEVA. The last one is defined only to $M=0.8$ M$_\odot$,
but it is reasonably close to our mass determinations to make an 
attempt at extrapolation. Because the BCAH98 set is not available for 
metallicities above solar, we show the isochrone for $Z=0.02$, but with
$Y = 0.282$ (normally 0.275 or 0.25), since it fits within errors to
our measurements of $M_{bol}$ for both systems. 

We consider the $M/M_{bol}$ as the most important figure because 
theoretical models seem to reproduce this combination of parameters 
properly for low-mass eclipsing binaries. For the ASAS-09 system 
we also gave more attention to the $M/(V-I)$ panel because 
in the distance determination we found 
$E(B-V)$ in this direction close to 0 and the uncertainty coming from 
dereddening is negligible. Our values of $V-I$ derived from
PHOEBE are nearly equal for both components, as expected for twin
stars. Thus they represent the color of the entire system. Our result
-- $1.17 \pm 0.13$~mag -- agrees well with 1.12 from \citet{szc08}.

We see that the bolometric luminosity of ASAS-09 is within errors in
agreement with every theoretical model presented. We also observe a 
situation typical for most of these systems -- the radii are larger and
temperatures lower than predicted. The isochrone that is 
closest either in $M/\log{T_{eff}}$ or $M/R$ plane is the BCAH98
for $Y=0.282$ and $t=1$ Gyr. This case fits our measurements best.
All models for $Z>0.02$ and $t=1$ Gyr agree within errors not only 
with the system's $M_{bol}$ and $V-I$, but also with $M_V$, which is 
not the case for models with $Z=0.02$. Thus we may suspect that
ASAS-09 is a main-sequence, metal-enriched and/or helium-depleted
system, probably nearly 1 Gyr old. We should also note that a lower 
metallicity results in a higher bolometric luminosity. 
Hence no $Z<0.02$ isochrone fits to ASAS-09 regardless of the age. 
This constitutes an argument against the
membership of ASAS-09 in the Pleiades moving group.

The situation is slightly different for ASAS-21. The absolute $V$ magnitude is
best reproduced in the $t=1$ Gyr, $Z=0.02$ case. However, in 
other $M/M_V$ panels the models still agree with the measurements.
The best $M_{bol}$ fit is obtained for the 5~Gyr case and the
worst one for metal-enriched models. On the other hand, $Z>0.02$
models reproduce temperatures but, as mentioned before,
it would be an unusual case for a low-mass eclipsing binary, 
and this panel is shown only for comparison. Thus ASAS-21 has a 
metal abundance similar to solar and is likely on the main sequence, 
but closer to 5 Gyr than to 1 Gyr, as suggested by the kinematics and 
the $M/M_{bol}$ plane. It is possible that ASAS-21 is a member of the 
Hercules Stream. The marginal discrepancy in the $V-I$ color 
(for two models only) may come from the uncertainty in $E(B-V)$ 
estimate and the system is slightly bluer. 

\section{Conclusions}
We presented the results of the analysis of two newly discovered 
spectroscopic double-lined low-mass eclipsing binaries -- 
ASAS~J093814-0104.4 and ASAS~J212954-5620.1 -- 
found by us through the spectroscopic survey for low-mass systems 
from the ASAS catalogue. We did not succeed in reaching the desired 3\% 
precision only in the mass of ASAS-09. However, we are reasonably sure
that for these two objects it is possible to reach a sub-\% level of 
precision in the absolute values of physical parameters with more 
spectroscopic and photometric data. We also estimate the scale of 
the influence of stellar activity on the parameters and find that 
this effect is most likely already compensated. The configuration 
of stellar spots in both systems changes on time scales of single weeks, 
which is a manifestation of a substantial level of activity 
-- their main characteristic. The evolution of spots is a known feature 
of active stars, but such a rapid one was not previously reported for 
low-mass eclipsing system (see Sect. \ref{sec_spots}) and rarely
reported in general. The variability of spots is a major obstacle 
in an analysis of binary systems such as ASAS-09 and ASAS-21.
In order to make the best-fitting parameters more precise, 
one should ideally make nearly simultaneous spectroscopic and 
photometric observations, at least for short period binaries like 
ASAS-09 and ASAS-21, to properly account for the influence of spots 
on the RV curve. 

As pointed out by \citet{win10}, in order to properly characterize and 
understand the properties of many active systems, several separate observing 
programs would be required. This would for example allow one to 
estimate the spread of measured radii and temperatures and possibly 
correlate it with the level of activity \citep{win10} and/or eventually 
evaluate how the derived radii and temperatures change with the 
configuration of spots, $H_\alpha$ emission, or other activity indicators. 
A more intensive monitoring of known active eclipsing binaries is also 
required to find more systems exhibiting an evolution of spots 
on various time scales.

Considering the above, we admit that parameters of ASAS-09 and 
ASAS-21 can be certainly improved and in our opinion these systems 
deserve more attention. We do not know many "stellar twins", like 
ASAS-09, among all known eclipsing systems, and untill now only three
other low-mass eclipsing systems were found. The high-precision
derivations of their parameters can serve as a critical test for
the validity of the Vogt theorem. This theorem says that stars of the 
same chemical composition have their radii, temperatures, and bolometric
luminosities strictly determined by their masses \citep{vog26}. 
Despite a wide (but controversial) criticism of this theorem, it is still
the basis of the stellar evolution theory and is taken for granted. 
No reliable empirical tests of this theorem have been made 
\citep[for a brief discussion and references see:][]{las02}.
Eclipsing "twins", very accurately characterized, should be perfect targets for
such a research. In the case of active systems the radii and temperatures
may be (and probably are) affected by the likely different level of activity 
of the components, but their $M_{bol}$ should not. A more extensive
study of these systems such as YY~Gem or ASAS-09 should validate the
Vogt theorem at least at a certain level.

Owing to its relatively low mass ratio, ASAS-21 is particularly 
interesting for testing evolutionary models. Systems with low $q$
provide much more stringent tests than binaries with nearly equal masses, 
and the parameters derived from the fitting of isochrones (if successful) 
are more accurate. The value of $q$ obtained for ASAS-21 is lower than 
for most of the known double-lined binaries \citep{luc06} and one of the 
lowest among late-type eclipsing systems. There is an additional difficulty 
in finding low-$q$ binaries, which is the low brightness ratio of the binary 
components. The identification of the faint lines of the secondary or the 
shallow secondary eclipse is then challenging and requires high SNR 
data \citep[for an extreme case of eclipse depths see][]{mac09}.
Considering that, any new object with $q\la0.85$ is valuable.


\begin{acknowledgements}
We would like to thank Hannah Worters, David Laney and John Menzies from the 
South African Astronomical Observatory for their support during our observations
at SAAO, Stephen Marsden and the Anglo-Australian Observatory 
astronomers for their help during our observing runs on the AAT.

This work is supported by the Foundation for Polish Science through a FOCUS grant 
and fellowship, by the Polish Ministry of Science and Higher Education through
grants N203 005 32/0449 and N203 3020 35. 
This research was co-financed by the European Social Fund and the national
budget of the Republic of Poland within the framework of the Integrated
Regional Operational Programme, Measure 2.6. Regional innovation
strategies and transfer of knowledge -- an individual project of the
Kuyavian-Pomeranian Voivodship "Scholarships for Ph.D. students
2008/2009 - IROP". KZ was supported by the Foundation for the Polish Science 
through grant MISTRZ and by the grant N N203 379936 from the Ministry of 
Science and Higher Education. This work is supported by the National Science 
Foundation through grants 0959447, 0836187, 0707634 and 0449001.
The observations on the AAT/UCLES have been funded by the Optical Infrared
Coordination network, a major international collaboration supported by the
Research Infrastructures Programme of the European Commissions Sixth
Framework Programme. 

This research has made use of the Simbad data base,
operated at CDS, Strasbourg, France. This publication makes use of data
products from the Two-Micron All Sky Survey (2MASS), which is a joint
project of the University of Massachusetts and the Infrared Processing and
Analysis Center/California Institute of Technology, funded by the National
Aeronautics and Space Administration and the National Science Foundation.

\end{acknowledgements}

\bibliographystyle{aa}

\end{document}